\documentstyle[aps,multicol,epsf,psfig]{revtex}

\begin{document}
\title{Optimal Quantum Clocks}
\author{
 V. Bu\v{z}ek$^{1,2}$, R. Derka$^{2}$, and S. Massar$^3$
}
\address{
$^{1}$ Institute of Physics, Slovak Academy of Sciences,
D\'ubravsk\'a cesta 9, 842 28 Bratislava, Slovakia
\\
$^{2}$ Faculty of Mathematics and Physics, 
Comenius University, Mlynsk\'{a} dolina F2, Bratislava, Slovakia
\\
$^3$ Service de Physique Th\'eorique, Universit\'e Libre de Bruxelles, 
CP 225, Bvd du Triomphe, B1050 Brussels, 
Belgium
}

\date{26 November 1998}
\maketitle
\begin{abstract}
A quantum clock must satisfy two basic constraints. The first is a
bound on the time resolution of the clock given by
the difference between its maximum and minimum energy
eigenvalues.
The second follows from Holevo's
bound on how much classical information can be encoded in a quantum
system. 
We show that asymptotically, as the dimension of the Hilbert space of
the clock tends to 
infinity, both constraints can be satisfied simultaneously.
The experimental
realization of such an optimal quantum clock using trapped ions is discussed.

\end{abstract}

\vspace{-0.7cm}
\begin{multicols}{2}

Recent technical advances in the
laser cooling and trapping of ions suggest that coherent manipulations
of trapped ions will be performed in the not too far future\cite{Wineland}. 
Apart from various important
applications such as quantum information processing or 
improving high-precision spectroscopy
these techniques also allow us to test
fundamental concepts of quantum theory. In particular, much deeper
insight into the problem of quantum measurement
can be obtained. 

In this Letter  we study the problem of 
building an optimal 
 quantum clock from an ensemble of $N$ ions. 
To be specific let us 
assume an ion trap with $N$ two-level ions
all in the ground state 
$|\Psi\rangle=|0\rangle\otimes\dots \otimes |0\rangle$.
This state is an eigenstate of the free Hamiltonian and thus cannot record
time. Therefore the first step in building a clock is to bring the
system to an initial state $\hat\Omega$ which is not an energy
eigenstate. For instance one can apply a Ramsey pulse whose shape and
duration is chosen such that it puts all the ions in the product state
\begin{eqnarray}
\hat{\Omega}_{prod}=\hat{\rho}\otimes\dots \otimes\hat{\rho}\quad ;\quad
\hat{\rho}=\frac{1}{2} (|0\rangle +| 1\rangle)(
\langle 0| + \langle 1|).
\label{2a}
\end{eqnarray}
We shall also consider more general states, but shall always take them
to belong to the symmetric subspace of the $N$ ions.
The basis vectors of this space will be denoted
$|m\rangle$, $m=0,1 \dots N$.
They are the completely symmetrized states of $N$ two-level
ions with $m$ ions in the ground state and $(N-m)$ ions in the excited 
state.
The states $|m\rangle$ have energy $E_m = m$ (this defines our unit
of energy, setting $\hbar =1$ then defines our unit of time).

The reason we can restrict ourselves to the symmetric subspace is that
we can map any clock state onto the symmetric subspace without
affecting its dynamics.
Indeed consider an initial state 
$\hat{\Omega} = |\psi\rangle \langle \psi |$
that does not belong to the symmetric
subspace of the atoms. We can decompose 
$|\psi \rangle = \sum_m \sum_\alpha c_{m\alpha} | m,\alpha
\rangle$ where $| m,\alpha
\rangle$, $\alpha = 1,..., ({ N \atop m})$ denote a basis of the
states with energy $ m $. Consider the unitary operator
$\hat{U}$ that maps the state $|\psi\rangle$ onto the symmetric subspace
without changing its energy:
$\hat{U} \sum_\alpha  c_{m\alpha} | m,\alpha
\rangle = c_m |m\rangle$ where $|m\rangle$ is as before the symmetric
state with energy $m$. Since $\hat{U}$ commutes with the
Hamiltonian the performance of the clock based on $|\psi\rangle$
is identical to the clock 
whose initial state is
 the symmetric state $|\psi_{sym} \rangle = \hat{U} | \psi
\rangle$.

After the preparation stage,
the ions evolve in time according to the Hamiltonian evolution
$\hat \Omega(t) = \hat U(t) \hat \Omega \hat U^\dagger(t)$, 
$\hat U(t)=\exp\{-i t
\hat{H}\}$.
The task is to determine the elapsed time $t$ by carrying out a
measurement on the ions.
Note that because of the 
indeterminism of quantum mechanics
it is impossible, given a {\it single} set of $N$ two-level ions, to
determine the elapsed time with certainty.
The best we can do is to 
{\it estimate} the elapsed time based on the result of a measurement
on the system\cite{Holevo}.

Making a good quantum clock requires a double
optimization. First of all one can optimize the measurement. This
aspect has been studied in detail in \cite{Holevo} where the best
measuring strategy was derived. But one can also optimize the initial
state $\hat \Omega$ of the system. It is this second optimization that is
studied in this Letter.

Before turning to the problem of optimizing the initial state
$\hat\Omega$, it is instructive to review 
the fundamental limitations on the performance of quantum
clocks.
Let us  first consider a simple
classical clock that can then be generalized to the quantum case. 
Our classical clock consists of a set of $n$ registers. Each
register is either in the $0$ or the $1$ state. Thus the classical
clock consists of $n$ bits, and can be in $2^n$ different states.
The dynamics of the clock is as follows: the first register flips from
$0$ to $1$ or from $1$ to $0$ every $2 \pi 2^{-n} $, the second register flips
every $ 2 \pi 2^{-n+1}$, etc... The last register flips every
$\pi$.
This clock thus measures time modulo $ 2 \pi$.
Note that this clock has an inherent uncertainty since it cannot
measure time with a precision better than $2 \pi 2^{-n}$. 
Throughout this Letter  the time uncertainty is defined 
as
\begin{equation}
\Delta t^2 = 
\overline{
\left[ (t_{estimate} - t_{true})({\rm mod}\ 2\pi) \right]^2
}\ .
\label{est}
\end{equation}
For the classical clock 
$\Delta t_{class} = { \pi 2^{-n} / \sqrt{3}}$.

It is straightforward to replace the 
classical clock by a quantum version. The
quantum clock consists of $n$ two level systems (qubits). The
first qubit has an energy
splitting between the levels of $2^{n-1}$ so that it has the
same period as the corresponding classical register. The second 
qubit has an energy
splitting of $2^{n-2}$ and so on up
to the last qubit that has an energy splitting of $1$.
Considered together these $n$ qubits constitute a quantum
system with $2^n$ equally spaced energy levels. 

The mapping between the Hilbert space of this abstract quantum 
clock and the
symmetric subspace of $N$ 2-level ions is straightforward when
$N+1 = 2^n$. Indeed in this case the dimension and energy spectrum of
both Hilbert spaces coincide. Note however that this is not a mapping
between the qubits of the clock and the ions individually, but between energy
eigenstates. 
This comparison between classical and quantum clocks suggests
that a quantum clock built out of $N$ ions
cannot behave better then a classical clock built out of $\ln(N+1)$
registers.
That this is indeed the case follows from two fundamental constraints:

The first constraint is a bound on the time resolution of the clock that
results from its energy spectrum.
Indeed the
time-energy 
uncertainty\cite{Uncertainty1,Uncertainty2}
\begin{eqnarray}
\Delta t  \Delta E \geq \frac{1}{2},
\label{Uncert}
\end{eqnarray}
where
$\Delta E^2 = {\rm Tr} (\hat H^2 \hat \Omega) - [{\rm Tr} (\hat H \hat
\Omega)]^2$ relates the uncertainty in the estimated time 
[defined by Eq. (\ref{est})] to the
spread in energy of the clock. In the present case there
is a state with a maximum energy uncertainty, namely
$|\psi_+ \rangle={1 \over \sqrt{2}} (|N\rangle +
|0\rangle )$ for which $\Delta E = N/2$. Inserting in Eq~(\ref{Uncert}) 
shows that for any clock
built out of $N$ ions
\begin{equation}
\Delta t \geq {1\over N}. 
\label{DeltaE} 
\end{equation}
Note that
one cannot attain equality in  Eq.~(\ref{Uncert}). Indeed 
the state $|\psi_+ \rangle$ evolves with a period $2 \pi /N$,
hence it necessarily has a large time uncertainty. Thus 
Eq.~(\ref{DeltaE}) is only a lower bound
and maximising the energy uncertainty as in \cite{Bollinger} is not
necessarily an optimal procedure.

The second fundamental constraint the clock must obey is a bound on
the total information it can carry. Indeed Holevo\cite{Holevo2} 
has shown that one
cannot encode and subsequently retrieve reliably
more than $n$ bits of classical information into $n$ qubits.
Letting a clock evolve for a given time interval $t$ can be viewed as
trying to encode information about the classical variable $t$ into the
state of the clock. Hence a measurement on our model clock (consisting
of $n= \ln (N+1)$ qubits) cannot retrieve 
more than $\ln (N+1)$ bits of
information about $t$.

Together these two bounds imply that the quantum clock cannot perform
better than the corresponding classical clock: it cannot carry more
information and it cannot have better resolution. 
But is it possible,
 by making an optimal measurement and choosing
in an astute manner the  initial state of the clock, to
make a quantum clock with similar performances  to the classical one?
Our main result is to show that this is indeed the case for clocks
built out of symmetric states of $N$ ions and to provide
an algorithm for constructing such an optimal clock.

The problem of constructing quantum clocks has been considered
previously in \cite{Salecker,Peres2}. 
However 
the best     clocks considered
in these papers are based on the phase state $|\Psi_0\rangle$
described below. As we shall see for these clocks the time
uncertainty is very large $\Delta t \simeq {1 \over \sqrt{N}}$
and is very far from reaching equality in Eq.~(\ref{DeltaE}).
Recently Vaidman and Belkind\cite{Vaidman} 
considered the problem of
clock for which equality holds in Eq.~(\ref{Uncert}). They showed that
in the limit of large $N$ the product states satisfy this
condition. However for the product state the energy uncertainty 
is very large $\Delta E = \sqrt{N}/2$ hence
they also do not saturate Eq.~(\ref{DeltaE}). Furthermore clocks based
on product state also do not attain Holevo's bound.

A similar approach to the one used here, namely optimizing both the
initial state and the measurement on a system of $N$ ions was considered
in \cite{Huelga} with the aim of using the ions as an improved frequency
standard. This problem can be rephrased 
in the following way: one disposes of a classical but noisy
clock which provides some {\it a priori} knowledge about
the time $t$ 
and one wants to improve the knowledge of $t$ by using the $N$ ions. 
On the other hand in the present Letter we suppose that there is
no prior knowledge about $t$.  
The other difference with the present work is that 
our aim is to study the
fundamental structure of quantum mechanics. We therefore
neglect the effect of noise during the preparation and
measurement stages and decoherence during the evolution. 
On the other hand taking these effects into account was 
central to \cite{Huelga}.

In order to find how to build an optimal 
clock we must delve in detail
into the functioning of this device.
We first recall Holevo's
results concerning the optimal measurement strategy\cite{Holevo}. The
measurement is described by a positive operator measurement (POVM), that is 
 a set $\{\hat O_r\}_{r=1}^ R$ of positive Hermitian operators such
that $\sum_r \hat O_r = \hat{1}$.
To each outcome $r$ of the measurement we associate an estimate $t_r$
of the time elapsed. The difference between the estimated time $t_r$
and the true time $t$ is measured by a cost function $f(t_r - t)$.
Here we note that because of the periodicity of the clock, $f$ has to be
periodic. We also take $f(t)$ to be an even function. 
The task is to minimize the mean value of the cost function
\begin{eqnarray}
\bar f = \sum_r \int_0^{2\pi} 
{\bf Tr}[\hat O_r\ \hat \Omega(t) ] f(t_r-t) \frac{d t}{2\pi}. 
\label{3}
\end{eqnarray}
To proceed we expand the cost function in Fourier series:
\begin{eqnarray}
f(t) = w_0 - \sum_{k=1}^\infty w_k \cos k t .
\label{cost}
\end{eqnarray}
The essential hypothesis made by Holevo is positivity of the Fourier
coefficients: $w_k \geq 0$, $(k=1,2,...) $.
He also supposes that the initial state $\hat\Omega =
|\omega\rangle\langle\omega|$, $|\omega\rangle = 
\sum_m a_m |m\rangle$  is a pure state (and
we make a phase convention such that $a_m$ is real and positive). 
He then shows that 
\begin{eqnarray}
\bar f \geq w_0 -{1 \over 2} \sum_{k=1}^\infty w_k \sum_{m,m' \atop
|m-m'|=k}
a_m a_{m'}, 
\label{meancost}
\end{eqnarray}
and equality is attained only if the measurement is of the form
\begin{eqnarray}
\hat O_r &=& p_r |\Psi_r\rangle\langle\Psi_r|\quad ; \quad 
p_r \geq 0\quad ; \quad
|\Psi_r\rangle = e^{ i t_r  \hat H} |\Psi_0\rangle ,\nonumber\\
|\Psi_0\rangle &=& {1 \over \sqrt{N+1} } \sum_{m=0}^N |m\rangle,
\label{measurement}
\end{eqnarray}
with the completeness relation $\sum_{r} \hat O_r 
= \hat 1$.

Several remarks about this result are called for:
\begin{enumerate}
\item Holevo supposed that the initial state is a pure state. If the
initial state is mixed, $\hat\Omega = \sum_i p_i |\psi_i\rangle \langle
\psi_i |$, then one finds that the corresponding cost is bounded by
the average of the bounds Eq.~(\ref{meancost}). This shows that in
building a good quantum clock one should always take the initial state
to be pure.

\item
Holevo considered covariant measurements in which the times $t_r$
takes the continuum of values between $0$ and $2 \pi$. But as shown in
\cite{Derka} the completeness relation can also be satisfied by taking
a discreet set of times $t_j = {2 \pi j \over N +1}$,
$j=0,...,N$. These ``phase''  
states 
$|\Psi_j\rangle$ \cite{Pegg} form an orthonormal basis of the
Hilbert space, and this measurement is therefore a von Neumann
measurement. It means that
it is not necessary to use an ancilla to make the optimal measurement
in this case.

\item 
In Eq.~(\ref{meancost}) only the values $k=0,...,N$ intervene
because of the condition $|m-m'|=k$. That is, only the first $N+1$
Fourier coefficients of the cost function are meaningful.

\item 
Because of the condition of positivity of $w_k$ not all cost
functions are covered by this result, but several important examples
are: 
$4\sin^2 {t \over 2} =2 (1 - \cos t )$,
$|t  ({\rm mod}\ 2 \pi)|$, $|\sin{t \over 2}|$, 
$- \delta[t ({\rm mod}\ 2 \pi)]$.
The most notable absent from this list is the quadratic deviation
$t^2$ [as defined in Eq.~(\ref{est})]
but it can be well approximated by the the first cost function since $
4 \sin^2 {t \over 2} \simeq t^2$ for $|t| << \pi$.
\end{enumerate}

We now turn back to the central problem of this Letter, namely how to
optimize the initial state of the system so that the time estimate is
as good as possible. This corresponds to minimizing the right hand side of
Eq.~(\ref{meancost}) with respect to $a_m$.
To this end let us define the matrix 
\begin{eqnarray}
F_{m m'} = w_0 \delta_{m, m'} - {1\over 2} \sum_{k=1}^N
w_k (\delta_{m, m'+k} + \delta_{m+k,m'}) .
\label{F}
\end{eqnarray}
We must then minimize $\bar f = a^T \hat{F} a$ under the
condition $a^T a =1$. Using a Lagrange multiplier we find the
eigenvalue equation $(\hat{F} - \lambda \hat 1) a =0$, and the
 task is therefore to find the smallest eigenvalue and eigenvector of
$\hat{F}$.

Let us first consider the cost function $f = 4 \sin^2 t/2$. The
advantage of this cost function is that the matrix $F$ is particularly
simple in this case. Furthermore for errors much smaller then $2\pi$,
$f$ and $\Delta t^2$ as defined in Eq.~(\ref{est}) coincide. Hence the
first constraint on clock resolution Eq.~(\ref{DeltaE})
can be approximately replaced by $\bar f  \geq 1/N^{2}$.

If the initial state is the product state 
Eq.~(\ref{2a}) then the mean cost 
is given by the expression
$
\bar f=2[ 1- 2^{-N} \sum_{i=0}^{N-1} \sqrt{ (\ ^N_i)(\ ^N_{i+1})}]
$ 
which for large $N$ decreases as
$\bar f \simeq 1/N$.
If the initial state is the phase state
$|\Psi_0\rangle$, 
then direct calculation shows that $\bar f = {2 \over N+1}$. Thus for
both states $\Delta t \simeq N^{-{1\over 2}}$ and one is very far from
attaining equality in Eq.~(\ref{DeltaE}).

However neither of these two states is optimal. To find the optimum we
note that the matrix $\hat F = 2 \delta_{m m'} -
\delta_{m m'+1} - \delta_{m+1 m'}$
can be viewed as the discretized second derivative operator
$\hat  F \simeq - {d^2 / dx^2}$ with von Neumann boundary conditions.
The lowest eigenvalue of $\hat F$ is therefore approximately
$\lambda_{min} \simeq {\pi ^2 / (N+1)^2}$ and the corresponding
eigenvector is 
\begin{eqnarray}
|\Psi_{opt}\rangle \simeq {\sqrt{2} \over \sqrt{N+1}}
\sum_m \sin{ \pi (m+1/2) \over N+1}
|m\rangle. 
\label{stateopt}
\end{eqnarray}
Thus in this case the 
cost decreases for large $N$ as
$\bar f_{opt} \simeq {\pi^2 \over ( N+ 1)^2} $
corresponding to 
$\Delta t_{opt} \simeq {\pi \over ( N+ 1)}$.
Therefore, up to a factor of $\pi$ the optimal clock attains the bound
Eq.~(\ref{DeltaE}).

It is also interesting to  
consider the situation where the cost function is the delta
function $f=-\delta[t({\rm mod}\ 2 \pi)]$. In
this case $F_{m m'} =-{1\over 2 \pi}$ for all $m,m'$. One
checks that 
the phase state  $|\Psi_0\rangle$ is the eigenvector of 
$\hat{F}$ with minimal
eigenvalue $\lambda = - {N \over 2 \pi}$.
 Note that one could also have taken a smeared delta function
since only the first $N+1$ terms intervene in Eq.~(\ref{meancost}). 
The smeared delta function is approximately zero everywhere except
in an interval of about $1/N$ around zero where it is equal to $N$. 
Thus for this cost function one wants to maximize the frequency with
which the
 estimated value of
$t$ is
within about $1 \over N$ of the true value. But there is no extra
cost if the estimated value is very far from the true one.
It is for this reason that taking $|\Psi_0\rangle$ as the 
initial state when the
cost function is $4\sin^2t/2$ is bad since  making
estimates that are wildly off is strongly penalized in that case.

\begin{figure}
\centerline {\psfig{width=7.0cm,file=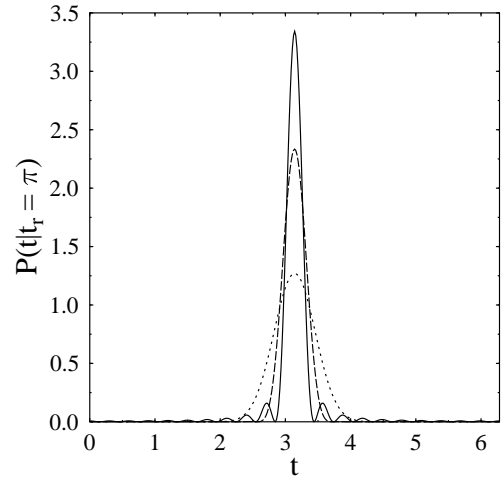}}
\begin{narrowtext}
\caption{ A plot of  
the {\it a posteriori} probability $P(t|t_r)$ that the time was $t$
given that the measurement yielded outcome $t_r=\pi$ for different
initial states ($N=20$). 
The dotted line corresponds to the product state
Eq.~(\ref{2a}), 
the solid line to the phase state $| \Psi_0\rangle$ (\ref{measurement}),
and the dashed line to the optimal state $| \Psi_{opt}\rangle$
(\ref{stateopt}).
}
\label{fig1}
\end{narrowtext}
\end{figure}

The mean value of a cost function gives only very partial information
about the sensitivity of a clock. The full information is encoded in
the probability $P(t|t_r) = P(\hat O_r | t ) P(t) /
P(\hat O_r)= P(\hat O_r | t ) {N+1 \over 2 \pi}$ 
that the true time is $t$ given that the
read out of the measurement is $t_r$.
In the case of the optimal state $|\Psi_{opt}\rangle$, one finds that
\begin{equation}
P_{opt}(t|t_r)
\simeq 
{\cal N}
{[1 + \cos(N+1)T](1 + \cos T) \over
\left[1 - \cos(T+{\pi \over N+1})\right]\left[1 - \cos(T-{\pi \over N+1})
\right]}, 
\end{equation}
where ${\cal N}\simeq{\pi\over 4(N+1)^3}$
and $T=t-t_r$.
This distribution (see Fig.~\ref{fig1})
has a central peak of width $3 \pi / (N+1)$ and
tails which decrease for $2\pi >> |t - t_r| >> N^{-1}$ as
$P_{opt}(t|t_r) \simeq N^{-3}|t  - t_r|^{-4}$. 

For the phase state $|\Psi_0\rangle$ one finds
\begin{equation}
P_{|\Psi_0\rangle}(t|t_r)= {1 \over 2\pi (N+1)}
{[1 - \cos(N+1)T] \over
(1 - \cos T)} .
\end{equation}
This distribution 
 has a slightly tighter central peak of width 
$2 \pi / (N+1)$ but the tails of the distribution decrease as
$N^{-1} |t  - t_r|^{-2}$. 
It is these slowly decreasing tails that give the main
contribution to $\Delta t\simeq N^{-{1\over 2}}$.

For the product state Eq.~(\ref{2a}) the distribution
$P_{prod}(t|t_r)$ has a very wide central peak of width $\simeq {1
\over \sqrt{N}}$. In this case it is the wide central peak that gives
rise to the large time uncertainty.

Using these distributions it is possible to calculate the number of
bits of information about time that is encoded in the outcomes of the
measurement. This is given by the mutual information
\begin{equation}
I = -\int dt P(t) \ln P(t) +
\sum_r P(t_r) \int dt P(t|t_r) \ln P(t|t_r) .
\nonumber
\end{equation}
In all cases the integral in the second term is dominated by the
central peak. Thus one finds that for the product state only ${1\over
2}\ln N$ bits of information about time are obtained, whereas for both
the states $|\Psi_0\rangle$ and $|\Psi_{opt}\rangle$ one obtains $\ln
N$ bits, thereby saturating Holevo's bound.

In summary we have seen that for different cost functions there are
different optimal clocks. There is of course no cost function that is
in an absolute sense better then another, and the choice of a
particular cost function depends on the physical context. Nevertheless
the experimental realization of a quantum clock based on the state 
$|\Psi_{opt}\rangle$ seems particularly desirable because it combines
the attractive features that the a posteriori probability
$P(t|t_r)$ has a tight central peak and rapidly decreasing tails.

Carrying out such an experiment with
trapped ions 
presents two main difficulties.
The first is the preparation
of the initial state $\hat\Omega$. 
Such coherent manipulation of trapped ions is one of the 
current experimental challenges.
A possibly important simplifying feature of the quantum clock is that 
since it is symmetric in
the $N$ ions one does not
need to address each ion individually. 
The second 
problem is to measure the phase states $|\Psi_j\rangle$.
Note that to perform such a measurement
it is sufficient to be able to realize coherent manipulations of the
ions
and to measure the 
projectors $|m\rangle\langle m|$, i.e. to measure 
the number of excited ions. Precisely 
this type of the measurement has recently been realized\cite{Turchette}. 
One may therefore hope that
for systems with moderate number of trapped ions it will be
feasible to make optimal quantum clocks in the not too far future.

Our special thanks to Lev Vaidman for his
criticism and careful rereading of the manuscript.
We also 
thank Win van Dam, Susana Huelga, Peter Knight and Christopher Monroe
for
helpful discussions. 
This work was in part supported by the Royal Society and by 
the Slovak Academy of Sciences.
V. B. and S. M. thank 
the Benasque Center for Physics where part of this work was
carried out. 
S.M. is a chercheur qualifi\'e du FNRS. He would like to
thank Utrecht University where most of this work was carried out.

\end{multicols}


\vspace{-0.5cm}
\begin{references}
\vspace{-1.5cm}

\bibitem{Wineland}
D.J. Wineland, {\it et al.}, { quant-ph 9710025} 


\bibitem{Holevo}
A.S. Holevo, {\it Probabilistic and Statistical
Aspects of Quantum Theory} (North-Holland, Amsterdam, 1982).

\bibitem{Uncertainty1} L. Mandelstam and I.E. Tamm, 
Izv. Akad. Nauk. USSR Ser. Fiz. {\bf 9}, No. 1-2, 122 (1945);
J. Phys. (Moscow) {\bf 9}, 249 (1945).

\bibitem{Uncertainty2}
S.L. Braunstein and C.M. Caves, Phys. Rev. Lett. {\bf 72}, 3439  (1994);
S.L. Braunstein, C.M. Caves, and G.J. Milburn,
Ann. Phys. {\bf 247}, 135 (1996).


\bibitem{Bollinger} Bollinger, et. al., Phys. Rev. A {\bf 54}, R4649 (1996)


\bibitem{Holevo2}  
A.S. Holevo, Probl. Peredachi Inf. {\bf 9}, 3 (1973)
[Probl. Inf. Transm. (USSR) {\bf 9}, 177 (1973)].



\bibitem{Salecker} 
H. Salecker and E.P. Wigner, Phys. Rev. {\bf 109}, 571 (1958).

\bibitem{Peres2} 
A. Peres, Am. J. Phys. {\bf 48}, 552 (1980).

\bibitem{Vaidman} L. Vaidman, private communication; O. Belkind,
{\it Quantum Stoppers}, M.Sc. Thesis, Tel Aviv University, 1998.


\bibitem{Huelga}
S.F. Huelga, {\it et al.},
{ Phys. Rev. Lett.} {\bf 79}, 3865 (1997).

\bibitem{Derka}
R. Derka, V. Bu\v zek, and A.K. Ekert, { Phys. Rev. Lett} {\bf 80},
1571 (1998).

\bibitem{Pegg}
D.T. Pegg and S.M. Barnett, { Europhys. Lett.} {\bf 6}, 483 (1988).

\bibitem{Turchette}
Q.A. Turchette, {\it et al.}, 
{ quant-ph 9806012}.

\end{references}
\end{document}